\documentclass[12pt]{report}

\usepackage[dvips]{graphicx}
\usepackage{amssymb}
\usepackage{amsmath}

\makeatletter
\renewcommand{\@oddhead}{\hfil \thepage}
\renewcommand{\@oddfoot}{}
\renewcommand{\thefootnote}{\alph{footnote}}

\makeatother

\begin{document}

\renewcommand{\bibname}{\large \textit{References}}

\begin{center}
\textbf{\large {Three Field Dynamics in (1+1)-dimensions}}
\end{center}
\begin{center}
\textbf{V.E. Kuzmichev} \footnote{$\ $ e-mail:
specrada@bitp.kiev.ua} \textbf{and V.V. Kuzmichev} \footnote{$\ $
e-mail: specrada@bitp.kiev.ua; vvkuzmichev@yahoo.com}
\end{center}
\begin{center}
\textit{Bogolyubov Institute for Theoretical Physics, National
Academy of Sciences of Ukraine, Metrolohichna St. 14b, Kiev, 03143
Ukraine}
\end{center}
\setcounter{footnote}{0}
\renewcommand{\thefootnote}{\arabic{footnote}}

\textbf{Abstract:} In a model of nonlinear system of three scalar
fields the problem on dynamics of a massive particle moving in
effective potential provided by two relativistic fields is
solving. The potentials for these fields are chosen in the form of
anti-Higgs and Higgs potentials. It is shown that the effective
potential has the shape of two-hump barrier localized in
spacetime. It tends to constant attractive potential at spacetime
infinity. The magnitude of this constant constituent is determined
by the Higgs condensate. It is shown that nonlinear equation of
motion of a particle has the solutions which describe the capture
of a particle by the barrier and the scattering on the
barrier.\\[0.5cm]

\textbf{1. Introduction.} Study of the dynamics of few fields
interacting between themselves is of interest in the connexion
with the problems of physics of nonlinear systems which arise in
quantum field theory \cite{1}, in plasma, in hydrodynamics, in
optics and so on \cite{2}. In QCD, for example, if one takes into
account spin, colour and flavour degrees of freedom in quark-gluon
system the requirement of gauge invariance of the theory leads to
clear structure of the Lagrangian \cite{3}. However the equations
of motion which follow from this Lagrangian prove to be too
complicated for analysis and solution. The modelling of peculiar
features of the Lagrangians of few-field systems with nonlinear
self-interaction and study on this basis of toy models allowing
exact or approximate solutions of the nonlinear equations of
motion seem to be important.

In the present article nonlinear system of three scalar fields in
(1+1)-dimensions is considered. One field is massive while the
potentials of two other fields are described by the anti-Higgs and
Higgs potentials. In the case when one can separate the slow and
rapid motion in spacetime the effective potential of interaction
of massive particle with ``force field'' provided by two
relativistic fields is obtained. The mechanism of the formation of
this potential is cleared up. It is shown that in general case it
has the shape of two-hump barrier localized in spacetime. The
potential tends to constant negative value at spacetime infinity.
This constant attractive interaction represents the energy
determined by the Higgs condensate. It arises in the system as a
result of reorganization of vacuum of the field with
self-interaction described by the Higgs potential. The equation of
motion of massive particle takes into account self-interaction of
its field and it is nonlinear. It is shown that the solutions of
this equation can describe both capture of a particle by the
barrier and the scattering on the barrier. The character of
solutions depends on coupling constants of relativistic fields.

\textbf{2. Effective potential.} We shall consider the nonlinear
system of three interacting scalar fields $\phi(x^{\mu}), \
\varphi_{1}(x^{\mu})$ and $\varphi_{2}(x^{\mu})$ in
two-dimensional spacetime $x^{\mu} = \{t, z\}$. Let us suppose
that the field $\phi$ is massive and complex and the fields
$\varphi_{i}$ are real. We define the Lorentz-invariant Lagrangian
in the form
\begin{equation}
\mathcal{L} = \partial_{\mu} \phi^{*}\,\partial^{\mu}\phi -
\phi^{*}m^{2}\phi +
\sum_{i=1}^{2}\left[\frac{1}{2}\,\partial_{\mu}\varphi_{i}\,\partial^{\mu}
\varphi_{i} - \phi^{*}U_{i}(\varphi_{i})\phi\right],
  \label{1}
\end{equation}
where $m$ is the mass of the field $\phi $. The functions
$U_{i}(\varphi_{i})$ are chosen in the form of anti-Higgs $(i =
1)$ and Higgs $(i = 2)$ potentials respectively
\begin{equation}
U_{i}(\varphi_{i}) = (-1)^{i}\left[-
\frac{\mu^{2}_{i}}{2}\,\varphi^{2}_{i} + \frac{\lambda_{i}}{4}\,
\varphi^{4}_{i}\right]
 \label{2}
\end{equation}
with coupling constants $\mu^{2}_{i} > 0$ and $\lambda_{i} > 0$.
Variational principle being applied to the Lagrangian (\ref{1})
leads to the set of three nonlinear field equations:
\begin{equation}
\partial^{2} \phi = - \left[\sum_{i} U_{i} + m^{2} \right]\phi,
  \label{3}
\end{equation}
\begin{equation}
\partial^{2} \varphi_{i} = - \left|\phi \right|^{2}\,\frac{\partial
U_{i}}{\partial \varphi_{i}}.
  \label{4}
\end{equation}
We shall limit ourselves to the consideration of the excitation
spectrum of the field $\phi$ with energies $E' \ll 2 m$. The
unitary transformation
\begin{equation}
\phi = \psi \exp(-\,i m t)
  \label{5}
\end{equation}
in mentioned nonrelativistic limit leads to the Schr\"{o}dinger
equation
\begin{equation}
i \, \partial_{t} \, \psi = \frac{1}{2 m}\left[-\,
\partial_{z}^{2} + V\right] \psi,
  \label{6}
\end{equation}
which describes the motion of a particle with the mass $m$ in the
effective potential
\begin{equation}
V = \sum_{i} U_{i}(\varphi_{i})
  \label{7}
\end{equation}
provided by the relativistic fields $\varphi_{i}$. The equations
(\ref{4}) - (\ref{7}) describe the system in which one can
separate two types of motion: slow variations of the
Schr\"{o}dinger field $\psi$ in the spacetime and rapid variations
of the fields $\varphi_{i}$. Neglecting the dependence of the
amplitude $|\phi| = |\psi|$ on variables $(t, z)$ we shall find
the solution of equations (\ref{4}) in the form of solitary waves
\cite{2,3}
\begin{equation}
\varphi_{1} = \pm \sqrt{\frac{2 \mu_{1}^{2}}{\lambda_{1}}}\
\mbox{sech} \left(\mu_{1} |\psi| s_{1} \right),
  \label{8}
\end{equation}
\begin{equation}
\varphi_{2} = \pm \sqrt{\frac{\mu_{2}^{2}}{\lambda_{2}}}\ \tanh
\left(\frac{\mu_{2}}{\sqrt{2}} |\psi| s_{2} \right),
  \label{9}
\end{equation}
where $s_{i} = (z - u_{i} t) / \sqrt{1 - u_{i}^{2}}$, and $u_{i}$
are the velocities (free parameters), $-1 < u_{i} < 1$.
Substitution of these solutions into (\ref{7}) gives the effective
potential
\begin{equation}
V(t, z) = \frac{\mu^{4}_{1}}{\lambda_{1}}\ \mbox{sech}
^{2}\left(\mu_{1} |\psi| s_{1}\right) \tanh ^{2} (\mu_{1} |\psi|
s_{1}) + \frac{\mu ^{4}_{2}}{4
\lambda_{2}}\left[\mbox{sech}^{4}\left(\frac{\mu_{2}}{\sqrt{2}}\,
|\psi| s_{2}\right) - 1 \right].
  \label{10}
\end{equation}
Its explicit form depends on the relations between the parameters
$\mu_{i}, \ \lambda_{i}$ and $u_{i}$. In Fig. 1 the typical
variants of the potential $V(t, z)$ in the stationary case $(u_{i}
= 0)$ as functions of $z$ are shown. The interactions which have
the shape of two-hump barrier are the most interesting. As it is
well known \cite{4} the scattering of a particle with positive
energy on such potential shows the resonant states. That
corresponds to the formation of quasistationary state and
subsequent quantum tunneling of a particle through the barrier
into the region of large values of $|z|$. The wave function of
such process out of range of barrier action has the form
\begin{equation}
\psi (z, t) = A(q) \left[\mbox{e}^{- i q z} - S(q)\,\mbox{e}^{i q
z} \right] \mbox{e}^{- i E t},
  \label{11}
\end{equation}
where $S(q)$ is the S-matrix, $q$ is the momentum of particle, $E$
is the energy, and the factor $A(q)$ is determined by the density
of the incident flux. Since the amplitude $|\psi|$ is the function
of $z$ which does not decrease on infinities the potential takes
the constant negative value in the limit $|\psi| s_{2} \rightarrow
\pm \infty$,
\begin{equation}
V_{c} = -\, \frac{\mu_{2}^{4}}{4\,\lambda_{2}}.
  \label{12}
\end{equation}
\begin{figure}
\begin{center}
\includegraphics[scale=0.9]{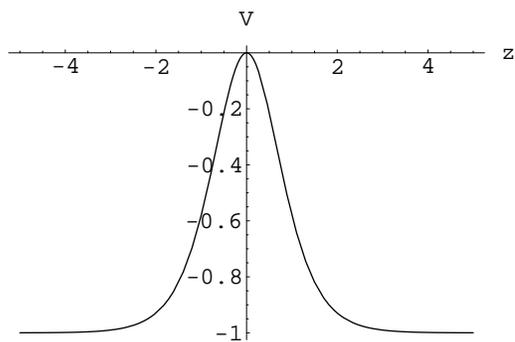}
\includegraphics[scale=0.9]{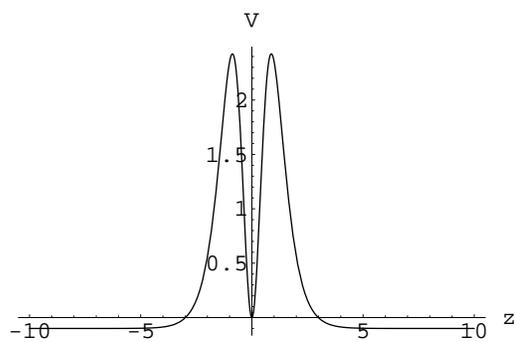}
\includegraphics[scale=0.9]{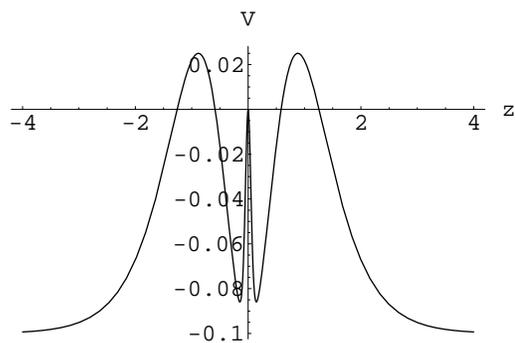}
\includegraphics[scale=0.9]{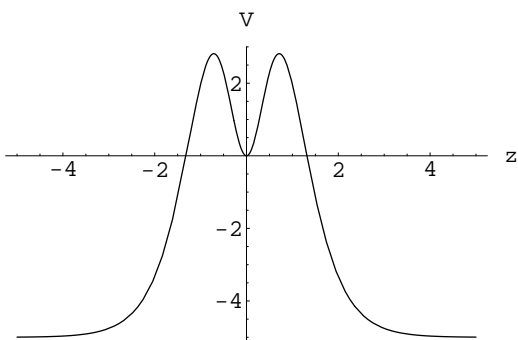}
\end{center}
\vspace{-2.cm} \caption{The effective potential (\ref{10}) in the
stationary case ($u_{i} = 0$) for different relations between the
parameters $\mu_{i}, \ \lambda_{i}$.}
\end{figure}
Hence the equation (\ref{6}) being nonlinear with respect to
unknown $\psi$ reduces to the ordinary Schr\"{o}dinger equation
with constant attractive potential in the shape of square step in
the limit of large values of $|z|$. If the energy $E = k^{2}/2m >
0$ counted from the value $V(0, 0) = 0$ then the momentum $q$ in
(\ref{11}) equals to $q = \sqrt{k^{2} - V_{c}}$. By means of
simple renormalization of energy, $\widetilde{E} = E - V_{c}/2m$,
the problem can be reduced to the problem of scattering on the
potential $\widetilde{V} = V - V_{c}$ which is short-range. We
note that such renormalization may be introduced already in the
initial Lagrangian (\ref{1}) by adding the constant term $- V_{c}$
to the potential $U_{2}$ which does not change the solitary wave
(\ref{9}). New potential $\widetilde{U}_{2}$ takes the form of the
Higgs potential
\begin{equation}
U_{2} \rightarrow \widetilde{U}_{2} = \frac{\lambda_{2}}{4}
\left[\varphi_{2}^{2} -
\frac{\mu_{2}^{2}}{\lambda_{2}}\right]^{2},
  \label{13}
\end{equation}
which is widely used in different applications of the field
theory. Such normalization of $U_{2}$ in particular is convenient
in cosmology during the study of the early Universe \cite{5, 6}.
For example, it can give the equation of state between the
pressure $p$ and the density $\epsilon$ in the form $p = -\,
\epsilon$ which leads to the exponential expansion of the Universe
in inflationary models \cite{7}.

In order to clarify the physical contents of the terms in
(\ref{10}) we rewrite it as
\begin{equation}
V(t, z) = \frac{1}{2\,|\psi|^{2}}\sum_{i=1}^{2}\left(1 -
u_{i}^{2}\right) \varepsilon_{i}(s_{i}) -
\frac{\mu_{2}^{4}}{4\,\lambda_{2}},
  \label{14}
\end{equation}
where $\varepsilon_{i}$ are the energy densities of the fields
$\varphi_{i}$ and $\varepsilon_{2}$ is renormalized according to
(\ref{13}). The functions $\varepsilon_{i}$ are localized in
spacetime and corresponding total energies $E_{i} =
\int_{-\infty}^{\infty}\!\! \varepsilon_{i}\,dz$ are finite. It
means that the fields (\ref{8}) and (\ref{9}) form the local
spacetime particle-like formations which are the carriers of the
effective interaction in the nonlinear three-field system. The
field $\varphi_{1}$ (\ref{8}) generates the part of interaction in
two-hump barrier shape and the field $\varphi_{2}$ (\ref{9})
smoothes it so that potential $V$ takes the value (\ref{12}) in
the limit of large $|z|$. This constant can be explained as a
result of spontaneous symmetry breaking \cite{8} of the Lagrangian
(\ref{1}) when the field $\varphi_{2}$ comes from unstable vacuum
state with $\varphi_{2} = 0$ to the state with stable vacuum
$\varphi_{2} = v$, where
\begin{equation}
v = \pm \frac{m_{H}}{2 \sqrt{\lambda}} = \pm
\frac{\overline{m}_{H}}{2 \sqrt{\overline{\lambda}}}
  \label{15}
\end{equation}
is the constant field (the Higgs condensate), $m_{H} = 2 \,
\mu_{2} \, m$ is the Higgs mass, $\lambda = \lambda_{2} \, m^{2}$
is Higgs self-constant corresponding to the free field
$\varphi_{2}$, $\overline{m}_{H} = m_{H}\,|\psi|$ and
$\overline{\lambda} = \lambda \,|\psi|^{2}$ are the effective
values of these quantities in the presence of the field $\psi$.
Taking (\ref{15}) into account the constant constituent (\ref{12})
can be expressed through the Higgs parameters,
\begin{equation}
V_{c} = - \frac{1}{64 \, \lambda} \, \frac{m_{H}^{4}}{m^{2}}.
  \label{16}
\end{equation}
This constant addition to the energy of interaction is proved to
be important during the calculation of the time delay caused by
the particle being located within the region inside the barrier
\cite{9}. This addition permits us introduce the natural cut off
of the confinement potential at distances of the order of the
range of barrier and provides suppression of the probability of
tunneling of a particle through the barrier of constant height and
width to the observed (or expected) value. The expression
(\ref{16}) itself can be considered as a relation between the
Higgs mass and the Higgs self-constant with the weight multiplier
$V_{c}$ being fixed independently \cite{9}.

\textbf{3. Localization of massive field.} According to (\ref{10})
and (\ref{13}) the renormalization potential $\widetilde{V}$ is
local in spacetime. Its form is determined by the sech-function
and the dependence of the $|\psi|$ on the variables $(t, z)$ does
not change the general behaviour of the effective potential.
Therefore the equation (\ref{6}) approximately can be considered
as the ordinary Schr\"{o}dinger equation. This approximation will
be good enough everywhere except the cases when
\begin{equation}
\mu_{1}^{2} |\psi|^{2} s_{1}^{2} \ll \frac{3}{5} \quad
\mbox{and/or} \quad \mu_{2}^{2} |\psi|^{2} s_{2}^{2} \ll
\frac{8}{3}.
  \label{17}
\end{equation}
In this region of values of $s_{i}^{2}$ the form of the potential
is determined by the wave function $\psi$ itself and the equation
(\ref{6}) takes the form of the nonlinear Schr\"{o}dinger equation
\begin{equation}
i\,\partial_{t} \psi + \frac{1}{2 m}\left[\partial_{z}^{2} +
\gamma \, |\psi|^{2} \right] \psi = 0
  \label{18}
\end{equation}
with the effective coupling constant
\begin{equation}
\gamma = \frac{\mu_{2}^{6}}{4 \lambda_{2}}\ s_{2}^{2} -
\frac{\mu_{1}^{6}}{\lambda_{1}}\ s_{1}^{2}.
  \label{19}
\end{equation}
If $\gamma > 0$ then the motion of a particle is determined mainly
by the part of the effective interaction formed by the
kink/antikink (\ref{9}) and the interaction is attractive. If
$\gamma < 0$ then the main contribution to $V$ is made by the
energy density of sech-wave (\ref{8}) which forms the barrier.
Since $s_{i}^{2}$ suppose to be small then their dependence on
spacetime variables in the equation (\ref{18}) can be neglected.
In such approximation it has single-soliton solution \cite{2}
\begin{equation}
\psi (z, t) = P \sqrt{\frac{2}{\gamma}}\ \mbox{sech}\left[P
\left(z - \frac{Q}{m}\,t\right)\right]\,\exp \left[i\, Q z + i\,
\frac{1}{2 m}\left(P^{2} - Q^{2}\right)t\right]
  \label{20}
\end{equation}
at $\gamma > 0$ and the boundary condition $\psi (z, t)
\rightarrow 0$ at $|z| \rightarrow \infty$. Here $P$ and $Q$ are
the free parameters (momenta). This solution describes the
traveling wave in the form of the local formation with the finite
total energy. The case $\gamma > 0$ is realized in the potentials
of the type shown in Fig. 1 (third plot). They have an attractive
part in the region inside the barrier. The function (\ref{20})
describes the particle which is captured by the potential well.
This phenomenon can be interpreted as capture of a particle by the
barrier with formation of long-live (quasistationary) state. The
particle can tunnel outside the barrier with small but finite
probability caused by long tails of the wave function (\ref{20})
coming from the sech-function.

This example shows that nonlinear character of motion of a
particle (i.e. a motion with self-interaction originated from the
interaction with the scalar fields $\varphi_{i}$) can lead to the
localization of its wave function in finite region of spacetime.

From the wave function (\ref{20}) one can restore the form of the
wave $\varphi_{2}$,
\begin{equation}
\varphi_{2} \sim \tanh \left[P \left(z -
\frac{Q}{m}\,t\right)\right].
  \label{21}
\end{equation}
Comparison with (\ref{9}) shows that the approximation (\ref{17})
describes the case of kink/antikink moving with the velocity
$u_{2} = (Q/m) \ll 1$ and the amplitude $|\psi|$ equals to the
effective value $\sqrt{2}P/\mu_{2}$.

At $\gamma < 0$ the particle gets to the repulsion region and
corresponding exact particular solution of the equation (\ref{18})
can be written in the form
\begin{equation}
\psi (z, t) = P \sqrt{\frac{2}{|\gamma|}}\,\sec \left[P \left(z -
\frac{Q}{m}\,t\right)\right]\,\exp \left[i\, Q z - i\, \frac{1}{2
m}\left(P^{2} + Q^{2}\right)t\right].
  \label{22}
\end{equation}
It describes the wave propagating in whole space with the energy
$E = (P^{2} + Q^{2})/2 m$ and amplitude which has the
singularities in the points where the function of secant turns
into infinity. These singularities apparently do not have the
physical meaning and are connected with the approximate character
of the calculations. The solution (\ref{22}) describes the
scattering of a particle on the barrier at small values of the
parameters $s_{i}^{2}$.

At $u_{1} = (Q/m) \ll 1$ the argument of secant is proportional to
$P s_{1} \sim |\psi| s_{1} \ll 1$ and we can write
\begin{equation*}
\psi (z, t) \approx P \sqrt{\frac{2}{|\gamma|}}\,\exp \left[i\, Q
z - i\, E\, t\right],
\end{equation*}
where the plane wave is distorted by the influence of the barrier
and the coefficient $P \sqrt{2/|\gamma|}$ is the effective
scattering amplitude.\\[0.5cm]

We should like to express our gratitude to Alexander von Humboldt
Foundation (Bonn, Germany) for the assistance during the research.


\begin{thebibliography}{99}
\itemsep -6pt plus 1pt minus 1pt
\thispagestyle{headings} \vspace{-1cm}
\bibitem{1} R.Rajaraman, Solitons and instantons (North-Holland
      Publishing Company, Amsterdam, New York, Oxford, 1982).
\bibitem{2} R.K.Dodd, J.C.Eilbeck, J.D.Gibbon, and H.C.Morris,
      Solitons and nonlinear wave equations (Academic Press, Inc, Ltd,
      London, 1982).
\bibitem{3} F.J.Yndurain, Quantum chromodynamics (Springer-Verlag,
      New York, Berlin, Heidelberg, Tokyo, 1983).
\bibitem{4} A.I.Baz', Ya.B.Zel'dovich, and A.M.Perelomov,
      Scattering, reactions, and decays in nonrelativistic quantum
      mechanics (Israel Program of Sci., Jerusalem, Transl., 1966).
\bibitem{5} A.D.Dolgov, Ya.B.Zeldovich, and M.V.Sazhin, Cosmology
      of the early Universe (Moscow University, Moscow, in Russian,
      1988).
\bibitem{6} V.V.Kuzmichev, Yad. Fiz. 60 (1997) 1707 [Phys. At.
      Nucl. 60 (1997) 1558].
\bibitem{7} A.D.Linde, Elementary particle physics and
      inflationary cosmology (Harwood, Chur, 1990).
\bibitem{8} Ta-Pei Cheng and Ling-Fong Li, Gauge theory of
      elementary particle physics (Clarendon Press, Oxford, 1984).
\bibitem{9} V.E.Kuzmichev and V.V.Kuzmichev, Confinement in a
      model with condensate of the scalar field (e-print, hep-th, April,
      2000).
\end{thebibliography}
\end{document}